\pdfoutput=1
\RequirePackage[l2tabu,orthodox]{nag}
\documentclass[a4paper,11pt]{article}
\usepackage[T1]{fontenc}
\usepackage[utf8]{inputenc}
\usepackage{pos}
\usepackage{microtype}
\usepackage{mathtools}
\usepackage{csquotes}
\usepackage{physics}
\usepackage{adjustbox}
\usepackage{slashed}
\usepackage{tikz}
\usepackage[outline]{contour}
\contourlength{2.5pt}
\usepackage{simpler-wick}
\usepackage{booktabs}
\usepackage{siunitx}

\usepackage[style=base]{caption}
\usepackage{subcaption}
\NewDocumentCommand\e{ m }{\mathrm{e}^{#1}}
\NewDocumentCommand\Op{}{\mathcal{O}}

\DeclareSIUnit\fm{\femto\metre}

\makeatletter
\DeclareFontFamily{OMX}{MnSymbolE}{}
\DeclareSymbolFont{MnLargeSymbols}{OMX}{MnSymbolE}{m}{n}
\SetSymbolFont{MnLargeSymbols}{bold}{OMX}{MnSymbolE}{b}{n}
\DeclareFontShape{OMX}{MnSymbolE}{m}{n}{
    <-6>  MnSymbolE5
   <6-7>  MnSymbolE6
   <7-8>  MnSymbolE7
   <8-9>  MnSymbolE8
   <9-10> MnSymbolE9
  <10-12> MnSymbolE10
  <12->   MnSymbolE12
}{}
\DeclareFontShape{OMX}{MnSymbolE}{b}{n}{
    <-6>  MnSymbolE-Bold5
   <6-7>  MnSymbolE-Bold6
   <7-8>  MnSymbolE-Bold7
   <8-9>  MnSymbolE-Bold8
   <9-10> MnSymbolE-Bold9
  <10-12> MnSymbolE-Bold10
  <12->   MnSymbolE-Bold12
}{}

\let\lAngle\@undefined
\let\rAngle\@undefined
\DeclareMathDelimiter{\lAngle}{\mathopen}%
                     {MnLargeSymbols}{'164}{MnLargeSymbols}{'164}
\DeclareMathDelimiter{\rAngle}{\mathclose}%
                     {MnLargeSymbols}{'171}{MnLargeSymbols}{'171}
\makeatother

\DeclareDocumentCommand\evsub{ s s m g }
{ 
	\IfNoValueTF{#4}
	{
		\IfBooleanTF{#1}
		{\vphantom{#3}\left\lAngle\smash{#3}\right\rAngle} 
		{\left\lAngle{#3}\right\rAngle} 
	}
	{
		\IfBooleanTF{#1}
		{
			\IfBooleanTF{#2}
			{\left\lAngle{#4}\middle\vert{#3}\middle\vert{#4}\right\rAngle} 
			{\vphantom{#3#4}\left\lAngle\smash{#4}\middle\vert\smash{#3}\middle\vert\smash{#4}\right\rAngle} 
		}
		{\vphantom{#3}\left\lAngle{#4}\middle\vert\smash{#3}\middle\vert{#4}\right\rAngle} 
	}
}

\usetikzlibrary{decorations.markings}
\usetikzlibrary{decorations.pathreplacing}
\usetikzlibrary{patterns}
\usetikzlibrary{fadings}
\usetikzlibrary {arrows.meta,positioning}

\tikzset{%
  ->-/.style={decoration={markings,mark=at position #1 with {\arrow{>}}},postaction={decorate}},
  ->-/.default=0.5
}

\title{Hadronic observables from master-field simulations}

\author*[a]{Marco Cè}
\author[b]{Mattia Bruno}
\author[c]{John Bulava}
\author[d]{Anthony Francis}
\author[e]{Patrick Fritzsch}
\author[c]{Jeremy R.\ Green}
\author[f]{Maxwell T.\ Hansen}
\author[g,h]{Antonio Rago}

\affiliation[a]{Albert Einstein Center for Fundamental Physics (AEC) and Institut für Theoretische Physik, Universität Bern, Sidlerstrasse 5, 3012 Bern, Switzerland}
\affiliation[b]{Dipartimento di Fisica \enquote{Giuseppe Occhialini}, Università degli Studi di Milano-Bicocca and INFN - Sezione di Milano Bicocca, Piazza della Scienza 3, 20126 Milan, Italy}
\affiliation[c]{Deutsches Elektronen-Synchrotron DESY, Platanenallee 6, 15738 Zeuthen, Germany}
\affiliation[d]{Institute of Physics, National Yang Ming Chiao Tung University, Hsinchu, Taiwan 30010}
\affiliation[e]{School of Mathematics and Hamilton Mathematics Institute, Trinity College Dublin, Dublin 2, Ireland}
\affiliation[f]{Higgs Centre for Theoretical Physics, School of Physics and Astronomy, The University of Edinburgh, Edinburgh EH9 3FD, United Kingdom}
\affiliation[g]{IMADA and CP3, University of Southern Denmark, Odense, Denmark}
\affiliation[h]{Department of Theoretical Physics, CERN, 1211 Geneva 23, Switzerland}

\emailAdd{marcoce@itp.unibe.ch}

\abstract{%
  Substantial progress has been made recently in the generation of master-field ensembles.
  This has to be paired with efficient techniques to compute observables on gauge field configurations with a large volume.
  Here we present the results of the computation of hadronic observables, including hadron masses and meson decay constants, on large-volume and master-field ensembles with physical volumes of up to $(18\,\mathrm{fm})^4$ and $m_\pi L$ up to $25$, simulated using $N_{\mathrm{f}}=2+1$ stabilized Wilson fermions.
  We obtain sub-percent determinations from single gauge configurations with the combined use of position-space techniques, volume averages and master-field error estimation.
}

\FullConference{%
The 39th International Symposium on Lattice Field Theory,\\
8th-13th August, 2022,\\
Rheinische Friedrich-Wilhelms-Universität Bonn, Bonn, Germany
}

\begin{document}

\makeatletter
\setlength\@tempdima{.78\paperheight}	
\divide\@tempdima\baselineskip		
\@tempcnta=\@tempdima			
\setlength\textheight{\@tempcnta\baselineskip}
\makeatother

\maketitle
\footskip    .02\paperheight	

\enlargethispage*{\baselineskip}
\section{Introduction}

Gauge-field configurations in a lattice theory with a mass gap have the \emph{stochastic locality} property, that is, gauge-invariant local fields at large physical separations are stochastically independent.
The master-field paradigm introduced by Lüscher~\cite{Luscher:2017cjh} proposes to use stochastic locality to obtain observable estimates from a single or at most a few representative gauge-field configurations on very large lattices, making use of the invariance under translations of the theory and of volume averages.

As a first application of this paradigm, stochastic locality has been used to compute the topological susceptibility at $T>T_c$ in master-field simulations of $\mathrm{SU}(3)$ Yang--Mills theory~\cite{Giusti:2018cmp}.
In a theory with fermions such as QCD, numerical simulations are performed after integrating out fermions exactly.
Hadronic observables in QCD are expressed in terms of contractions of quark propagators whose locality is not manifest.
Moreover, the sheer size of the lattices requires stabilising measures that have been studied in ref.~\cite{Francis:2019muy}.
These include a slight modification of the standard $\order{a}$-improved lattice Dirac operator, replacing the HMC with the stochastic molecular dynamics (SMD) algorithm, employing quadruple-precision lattice sums and uniform-norm stopping criteria for the Dirac equation solver.
Recent progress in master-field simulations has been presented at the Lattice 2021 conference~\cite{Fritzsch:2021klm,Ce:2021akh} and at this conference~\cite{Fritzsch:2022lattice}.
In these proceedings we further develop the position-space techniques introduced in ref.~\cite{Ce:2021akh}, by presenting an estimator for position-space correlators that scales efficiently with the volume.

Estimation of observables on master fields is explained in details in ref.~\cite{Luscher:2017cjh}.
In summary, the expectation value $\ev{\Op(x)}$ of a local field $\Op(x)$ is obtained averaging over translations
\begin{equation}
  \evsub{\Op(x)} = \frac{1}{V} \sum_z \Op(x+z) , \qquad \ev{\Op(x)} = \evsub{\Op(x)} + \order{V^{-1/2}} ,
\end{equation}
with the variance of this estimator given by
\begin{multline}
\label{eq:mf_var_O}
  \sigma^2_{\evsub{\Op}}(x) = \ev{ [\evsub{\Op(x)} -\ev{\Op(x)}]^2 }
                          = \frac{1}{V} \sum_{y} \ev{\Op(y) \Op(0)}_c \\
                          = \frac{1}{V} \left[ \sum_{\abs{y}\leq R} \ev{\Op(y) \Op(0)}_c + \order{\e{-mR}} \right]
                          = \frac{1}{V} \left[ \sum_{\abs{y}\leq R} \evsub{\Op(y) \Op(0)}_c + \order{\e{-mR}} + \order{V^{-1/2}} \right] ,
\end{multline}
where in the second line we first used the fact that the connected correlator of the local field $\Op(x)$ decays exponentially with spacetime separation, and then we applied again translation averages.

\section{Position-space correlators}
\label{sec:postion_space_correlators}

In this work we focus on correlation functions in position space
\begin{subequations}\label{eq:position_space_corr}
\begin{gather}
  C_{PP}(x) \rightarrow \frac{c_P^2}{4\pi^2} \frac{m_\pi}{|x|} K_1(m_\pi|x|) , \label{eq:cPP} \\
  C_{AP,\mu}(x) \rightarrow \frac{c_A c_P}{4\pi^2} \frac{x_\mu}{|x|} \frac{m_\pi}{|x|} K_2(m_\pi|x|) ,\\
  C_{AA,\mu\nu}(x) \rightarrow \frac{c_A^2}{4\pi^2} \left[ -\delta_{\mu\nu} \frac{1}{x^2} K_2(m_\pi|x|) + \frac{x_\mu x_\nu}{x^2} \left( \frac{m_\pi}{|x|} K_1(m_\pi|x|) + \frac{4}{x^2} K_2(m_\pi|x|) \right) \right] ,\\
  C_{NN}(x) \rightarrow \frac{c_N^2}{4\pi^2} \frac{m_N^2}{|x|}\left[ K_1(m_N|x|) + \frac{\slashed{x}}{|x|} K_2(m_N|x|) \right] , \label{eq:cNN}
\end{gather}
\end{subequations}
where the subscript indicates the two-point function of either pseudoscalar densities $P=\bar{u}\gamma_5 d$, axial current $A_\mu=\bar{u}\gamma_\mu\gamma_5 d$ or nucleon spinor $N=\epsilon_{abc}(u_a^TC\gamma_5d_b)u_c$, as a function of the source-sink separation $x$.

In eqs.~\eqref{eq:position_space_corr}, the asympotic behaviour for $x\to\infty$ of these correlators is given assuming the symmetries of the continuum theory in an infinite volume.
From position-space correlators one can extract simple hadronic observables, including the masses $m_\pi$ and $m_N$ and the decay constant $f_\pi=c_A/m_\pi$, as demonstrated in ref.~\cite{Ce:2021akh}.

Once computed on the lattice as discussed in the following section, these correlators as a function of the four-dimensional source-sink separation $x$ include lattice discretization effects that break the rotational symmetry and depend on the direction of $x$.
In this study, we limit ourselves to the radial correlators $\mathring{C}(r)$ introduced in ref.~\cite{Ce:2021akh} that are averaged over $S^3(r)=\{x\in\mathbb{R}^4:\abs{x}=r\}$, the $3$-sphere of radius $r$, and by construction depend only on the radial coordinate $r=\abs{x}$.
While $\mathring{C}_{PP}(r)=C_{PP}(x)$, for the $AP$-correlator $C_{AP,\mu}(x)$ we contract the open $\mu$ index with the only available four-vector $x_\mu$ to obtain a scalar, $\mathring{C}_{AP}(r) = x_\mu C_{AP,\mu}(x) \to \frac{c_A c_P}{4\pi^2} m_\pi K_2(m_\pi r)$.
In the case of $C_{AA,\mu\nu}$ there are two ways to obtain a scalar,
\begin{gather}
  \mathring{C}_{AA}^{(1)}(r) = \delta_{\mu\nu} C_{AA,\mu\nu}(x) ,\qquad 
  \mathring{C}_{AA}^{(2)}(r) = x_\mu x_\nu C_{AA,\mu\nu}(x) , 
\end{gather}
and similarly for the nucleon correlator that is a spinor, with $\slashed{x}=\gamma_\mu x_\mu$,
\begin{equation}
  \label{eq:cNN_ring}
  \mathring{C}_{NN}^{(1)}(r) \equiv \tr C_{NN}(x) ,\qquad 
  \mathring{C}_{NN}^{(2)}(r) \equiv \tr\slashed{x}C_{NN}(x) . 
\end{equation}
On the lattice, an estimator of these radial correlators is given by
\begin{equation}
  \mathring{C}(r) = \frac{1}{\mathrm{r}_4(r^2)} \sum_{\abs{x}=r} C(x)
\end{equation}
where $\mathrm{r}_4$ is defined in ref.~\cite{Ce:2021akh}.

We note that the symmetry of $S^3(r)$ is broken not only by $a\neq 0$ but also by the finite size of the hypercubic box and by the fact that we choose antiperiodic (instead of periodic) boundary conditions in one of the four dimensions for quarks.
However, as we show in section~\ref{sec:results} these boundary effects are not visible at the current level of precision on the master-field lattices in table~\ref{tab:parameters} considered here, differently from what we observed on smaller volumes~\cite{Ce:2021akh}.

\enlargethispage*{\baselineskip}
\section{Grid of point sources estimator}
\label{sec:grid_estimator}

\begin{table}[tb]
  \caption{Parameters of the master-field lattices used in this study (with $a\approx\SI{0.094}{\fm}$ and $m_\pi\approx\SI{270}{\MeV}$, see also ref.~\cite{Fritzsch:2021klm}), together with information on the statistics used in the observable computation as explained in section~\ref{sec:grid_estimator}.}
  \label{tab:parameters}
  \centering
  \begin{tabular}{lS[table-format=3]S[table-format=2]S[table-format=2.1]S[table-format=1]S[table-format=2]S[table-format=3]S[table-format=3]S[table-format=2]S[table-format=4]}
    \toprule
      & {$L/a$} & {$L$ [\si{fm}]} & {$m_\pi L$} & {$n_{\text{cnfg}}$} & {$b/a$} & {$|G|$} & {$n_{\text{shift}}$} & {$b_{\text{shift}}/a$} & {$n_{\text{point}}$} \\
    \midrule
    A &    96   &          9      &      12.5   &              5      &    48   &     8   &            512       &              12        & 4096 \\
    B &   192   &         18      &      25     &              2      &    48   &   128   &             32       &              24        & 4096 \\
    \bottomrule
  \end{tabular}
\end{table}

The simplest way to compute the correlators introduced in section~\ref{sec:postion_space_correlators} numerically is to solve the Dirac equation on a point source, that is, a source spinor that is supported on a single lattice point, and subsequently perform the suitable contractions of spinor and space-time indeces.
A consequence of this naive strategy applied on gauge-field configurations with a large volume is that the effort for each correlator point source scales proportionally with the volume, which is clearly not optimal.
Indeed, most of the resources are spent in computing the correlator at a distance from the source of multiple correlation lengths, which has an exponentially suppressed contribution to the physics and in most of the cases is completely dominated by noise.

Instead, we would like to exploit stochastic locality to define estimators that scale efficiently with the volume and are suitable for master-field applications.
Taking as an example the radial correlators $\mathring{C}(r)$ introduced in section~\ref{sec:postion_space_correlators}, let us assume that we are interested in physics that can be extracted from correlators up to a maximum radial source-sink separation $r_{\text{max}}$.
Ref.~\cite{Luscher:2017cjh} sketches a decomposition of the lattice in space-time domains, or blocks, that are physically large, such that all the lattice points within an $r_{\text{max}}$ distance from a source point at the centre of each block are within the same block.
This implies a block size $b>2r_{\text{max}}$.
Solving the Dirac equation in each block, imposing Dirichlet boundary conditions at the block boundary of the gauge field, one can decouple the computational cost of the estimator from the volume of the global lattice.
However, this method introduces boundary effects that can be large for sink points close to the boundaries~\cite{Luscher:2017cjh,Ce:2016idq}, see also refs.~\cite{Giusti:2022xdh,Saccardi:2022lzd}.
We leave the exploration of this direction for future work, and we focus here on a simpler approach that does not require a dedicated correction computation.

We introduce a set of lattice points $G$ that are separated (on average) by a physical distance constant in the volume, such that the number of points $|G|\propto V$, that is, it grows proportionally with the volume.
On these point we introduce stochastic sources that satisfy
\begin{equation}
\label{eq:noise_def}
  \ev{ \eta_i(x) \eta_j^\dagger(y) }_\eta = \delta_{ij} \delta_{xy} I \qquad \text{for $x,y\in G$} ,
\end{equation}
where $I$ is the identity matrix in spin and colour space.
By contracting at the sink with stochastic noise corresponding to each coordinate $y\in G$ one obtains $|G|\propto V$ samples of the quark propagator, one for each $y\in G$, from a single global-lattice inversion that is $\order{V}$ computationally.
Each sample has a spurious contribution of stochastic nature from source points $x\neq y$, which is suppressed by averaging the quark propagator over a number of sources $n_{\text{src}}$ and does not contribute to the expectation value.
Mesonic two-point functions that contract two quark propagators require $n_{\text{src}}\geq 2$ to obtain an unbiased estimator, that in the case of the pseudoscalar-density two-point function reads
\begin{equation}
\label{eq:grid_estimator}
  C_{PP}^G(x;y) =
  \frac{1}{n_{\text{src}}(n_{\text{src}}-1)} \sum_{i\neq j} \Re\left[
    \psi_i^\dagger(x+y)\psi_j(x+y)
    \eta_j^\dagger(y)\eta_i(y) \right],
\end{equation}
where $\psi_i(x)=\sum_y D^{-1}(x;y)\eta_i(y)$ and the double sum over $i\neq j$ can be computed in $\order{n_{\text{src}}}$ cost.

\begin{figure}[t]
  \centering
  \begin{tikzpicture}[scale=0.8]
    \clip (-1.5,-0.5) rectangle (16.5,6.5);
    \draw[thin,dotted] (-5,-5) grid (20,10);
    \draw[dashed] ( -5,+10) -- (+10, -5);
    \draw[dashed] (  0, -5) -- (+15,+10);
    \draw[dashed] ( +5,+10) -- (+20, -5);
    \draw[thin,dashed] (5,5) circle [radius=3.535533906];
    \draw[thin,<->] (-0.5,0) -- (-0.5,5) node[midway] {\contour{white}{$b$}};
    \draw[thin,<->] (5,5) -- (7.5,2.5) node[midway,right] {\contour{white}{$r_{\text{max}}=\sqrt{2}b/2$}};
    \draw[thick,->-]           ( 5,5) to [bend right]    ( 4,2);
    \draw[thick,->-]           ( 4,2) to [bend right]    ( 5,5);
    \draw[thick,->-,lightgray] (10,0) to [bend right=20] ( 4,2);
    \draw[thick,->-,lightgray] ( 4,2) to [bend right=20] (10,0);
    \fill ( 0,0) circle [radius=3pt];
    \fill ( 5,5) circle [radius=3pt] node[above] {\contour{white}{$y\in G$}};
    \fill (10,0) circle [radius=3pt];
    \fill (15,5) circle [radius=3pt];  
    \draw[fill=white] ( 4,2) circle [radius=3pt] node[below left] {\contour{white}{$x$}};
  \end{tikzpicture}
  \caption{%
    Sketch of the estimator with a grid of point sources over a two-dimensional window of the lattice.
    The set $G$ of source points $\protect\tikz{\fill (0,0) circle(3pt);}\in G$ is a regular grid with spacing $b$ and even point only.
    A mesonic two-point function is evaluated at sink point $x$ that is in the domain defined by $y\in G$ and within a distance $r_{\text{max}}$ from $y$.
    One of the spurious contributions from the \enquote{wrong} source is shown in light grey.
  }\label{fig:sources_sketch}
\end{figure}
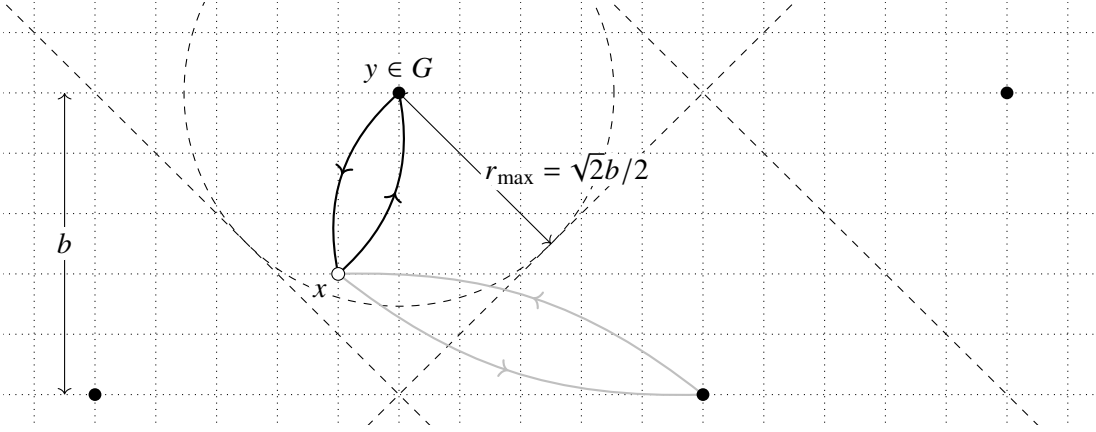

In this approach%
, since $|G|\propto V$, efficient scaling of the solutions of the Dirac equation is achieved.
Moreover, $G$ implicitly realises a domain decomposition by labelling each lattice point with the closest $y\in G$.\footnote{
  Up to points equidistant from two or more $y\in G$ that require additional conditions to be assigned to a domain.
}
Eq.~\eqref{eq:grid_estimator} is in principle valid for any $x$ and $y$, but if only $(x,y)$ pairs that are in the same domain are considered then one can compute efficiently all the $|G|\propto V$ contributions with a single $\order{V}$ pass over the whole lattice.
This realises the optimal volume scaling for the contractions too.
It also lowers the required $n_{\text{src}}$ since the \enquote{correct} source $y$ is always the closest to the sink $x$ and spurious contributions are further suppressed by the longer source-sink separation.\footnote{%
These spurious contributions are only stochastic and do not modify the expectation value, although we note that they can have different quantum numbers and decay slower than the correlator being estimated.
}
Moreover, it implies $r_{\text{max}}=\min_{x,y\in G}\abs{x-y}/2$, that is, the minimum of the semidistance of points in $G$.
Therefore, $G$ has to be sparse enough for correlators at the relevant radial separations $r\leq r_{\text{max}}$ to be accessible.

We study this setup on two sets of a few master fields whose parameters are given in table~\ref{tab:parameters}.
The master fields in both sets are hypercubic boxes with equal extent in each dimension denoted by $L$, such that the volume is $V=L^4$.
The $L=192a$ master fields denoted by B ($n_{\text{cnfg}}=2$) have exactly 16 times, twice in each dimension, the volume of the ones with $L=96a$ in set A ($n_{\text{cnfg}}=5$) and otherwise identical parameters, and we can thus define equivalent $G$s on both sets and study the volume scaling.
We employ $\mathrm{U}(1)$ noise that satisfies eq.~\eqref{eq:noise_def}.
The simplest choice for $G$ is a regular grid with spacing $b$, which matches the domain decomposition proposed in ref.~\cite{Luscher:2017cjh}, with $b=48a$ being a suitable choice in our case. 
However, the definition of $G$ is more flexible.
In this work, we employ a grid with only even (or equivalently odd) points, which results in $r_{\text{max}}=\sqrt{2}b/2\simeq 33.94a$ instead of $b/2=24a$, at the cost of halving the number of points on the grid.\footnote{%
This results in a doubled $|G|r_{\text{max}}^4/V$ density.
Indeed, it corresponds to a $D_4$ lattice (or equivalently $F_4$ lattice) that has the densest known packing of equal spheres in four dimensions~\cite{Conway:1999packing}.
}
The total number of points is thus $|G|=(L/b)^4/2$ that evaluates to $8$ and $128$ for A and B respectively.
We fix $n_{\text{src}}=2$ and with the current precision we do not observe deviations from the expected behaviour, especially at $r$ close to $r_{\text{max}}$, that can be attributed to spurious contributions.
Further optimisation such as systematically and exactly removing the closer spurious contributions, e.g.\ with hierarchical probing~\cite{Stathopoulos:2013aci}, are not explored here.

The statistics obtained with a single source, e.g.\ eight points on each master field in A, is limited by the need of balancing the density of $G$ with a lower limit on the $r_{\text{max}}$ suitable to extract long-range physics.
To increase the statistics we simply propose to recompute eq.~\eqref{eq:grid_estimator} on $n_{\text{shift}}$ sources, each time shifting $G$ to have a distinct support.
This is done four times for each direction in the case of A and twice for each direction in B.
An extra factor of two is obtained by pairing each even-only $G$ with the corresponding odd-only, leading to $n_{\text{shift}}=512$ and $32$ for A and B respectively.
Combined with $|G|$, the final result is the same number of source points $n_{\text{point}}=4096$ for both volumes, on a regular grid with spacing $b_{\text{shift}}=12a$ and $24a$ for A and B respectively.
Ignoring that on the A lattices source points are on average twice as close and thus potentially more correlated than on B, in our setup we have same statistics for each gauge field configuration for both A and B.
Crucially, thanks to the optimal volume scaling of the stochastic grid correlator, this matching statistic has been obtained at an equivalent computational cost.

\enlargethispage*{\baselineskip}
\section{Master-field errors}
\label{sec:master_field_errors}

The estimator in section~\ref{sec:grid_estimator} applied to the radial correlator leads to a collection of up to $4096$ correlators for each master-field configuration on a regular grid of source points with spacing $b_{\text{shift}}=L/8$.
Applying stochastic locality, the expectation value $\ev{\mathring{C}(r)}$ is given up to volume-suppressed corrections by the translation average
\begin{equation}
  \ev{\mathring{C}(r)} = \evsub{\mathring{C}(r)} + \order{V^{-1/2}} = \frac{1}{V} \sum_{y\in G} \mathring{C}(r;y) + \order{V^{-1/2}}
\end{equation}
where the $y$ in $\mathring{C}(r;y)$ denotes the source point.
The error of this estimator can be estimated applying eq.~\eqref{eq:mf_var_O} with $\Op(y)=\mathring{C}(r;y)$
\begin{equation}
  \ev{ [\evsub{\mathring{C}(r)} -\ev{\mathring{C}(r)}]^2 } = \frac{1}{V} \left[ \sum_{\abs{y}\leq R} \evsub{\mathring{C}(r; y) \mathring{C}(r; 0)}_c + \order{\e{-mR}} + \order{V^{-1/2}} \right] ,
\end{equation}
where again the sum over the source coordinates $y$ is performed over the grid of point sources.

Finding the optimal $R$ to truncate the sum in the r.h.s.\ has a clear analogy with the well-known $\Gamma$ method introduced by Wolff to deal with autocorrelation in Monte Carlo time and estimate an error with less errors~\cite{Wolff:2003sm}, and leads to a generalisation of the Madras--Sokal formula for the statistical error of the error~\cite{Madras:1988ei,Ce:2023grabbag}.
This can be implemented in a resource efficient way by computing the correlation between grid points with higher-dimensional fast Fourier transforms.
The optimal $R$ depends on the observable.
In particular, since each value of the correlator radial source-sink separation $r$ defines a distinct observable with different spacetime support, $R$ is a function of $r$.

Alternatively, one can apply a four-dimensional binning of the point sources in the grid into blocks.
For instance, blocks of size $(24a)^4$ bin $16$ point sources on A and only one point source on B according to the spacing $b_{\text{shift}}$ in table~\ref{tab:parameters}, while blocks of size $(48a)^4$ bin $256$ and $16$ point sources respectively.
We tested these two bin sizes and observed that this leads to a stable error estimate.
In the following, we show results obtained in the more conservative case, that is, with blocks of size $(48a)^4$.

We note that master-field error estimation can be combined with standard methods based on an ensemble of gauge field configurations, e.g.\ with a five-dimensional variant of the $\Gamma$ method in spacetime coordinates and Monte Carlo time.
Explorations in this direction can be found in ref.~\cite{Lehner:2022edinburgh}.

\section{Numerical results}
\label{sec:results}

\begin{table}[bt]
  \caption{Numerical results for hadronic observable with errors estimated \emph{à la} master field.}
  \label{tab:results}
  \centering
  \begin{tabular}{lS[table-format=3]S[table-format=1.5(2)]S[table-format=1.3(1)]S[table-format=1.4(1)]}
    \toprule
      & {$L/a$} &  {$am_\pi$} & {$am_N$} & {$af_\pi^{\text{bare}}$} \\
    \midrule
    A &    96   & 0.12628(33) & 0.500(6) & 0.0890(3) \\
    B &   192   & 0.12601(19) & 0.487(8) & 0.0885(4) \\
    \bottomrule
  \end{tabular}
\end{table}

We computed $m_\pi$, $m_N$ and $f_\pi$ using position-space correlators on the sets of master fields whose parameters are listed in table~\ref{tab:parameters}.
The results for these hadronic observables are listed in table~\ref{tab:results}.

\begin{figure}[t]
  \centering
    \scalebox{.45}{\includegraphics{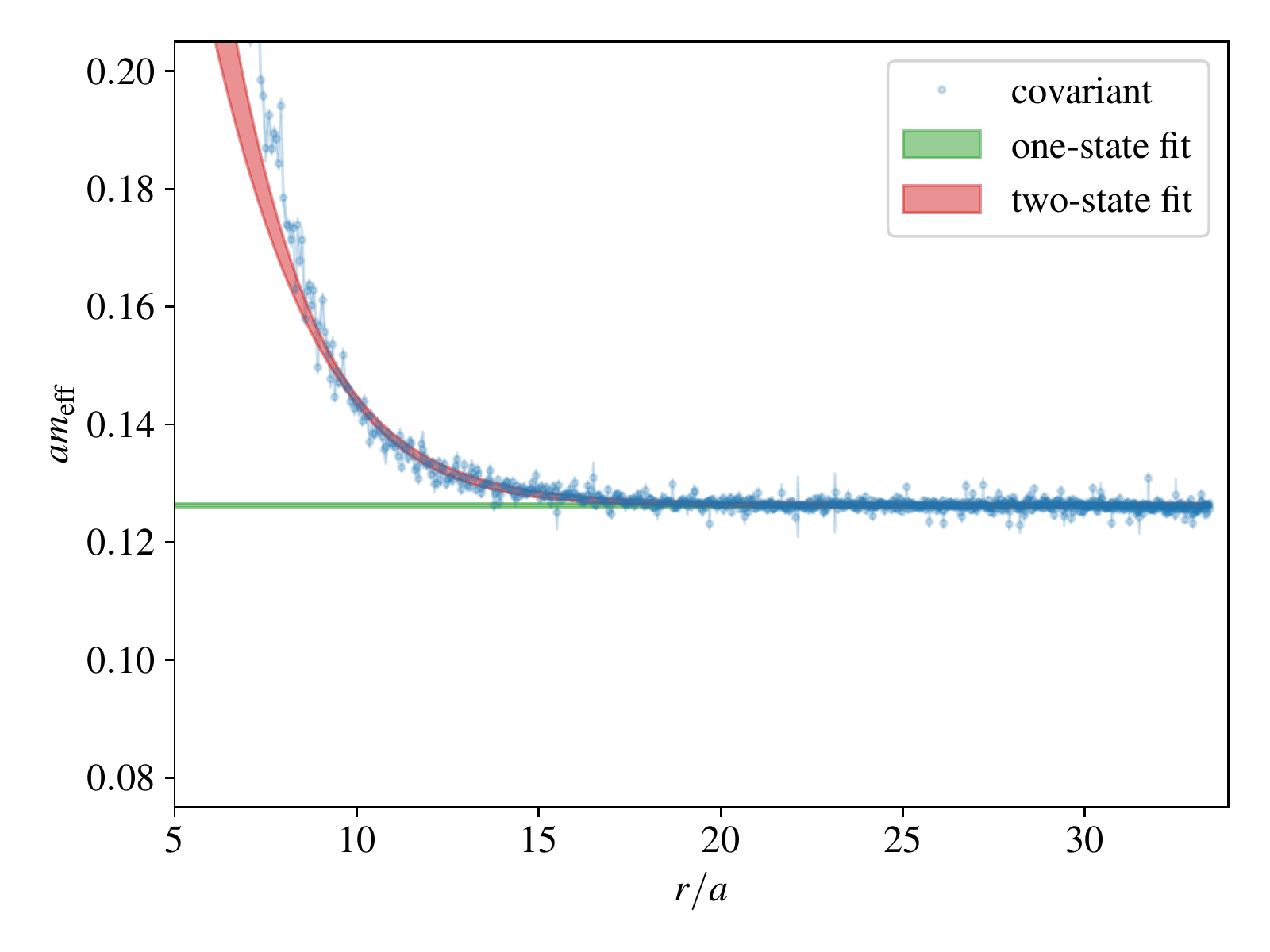}}%
    \scalebox{.45}{\includegraphics{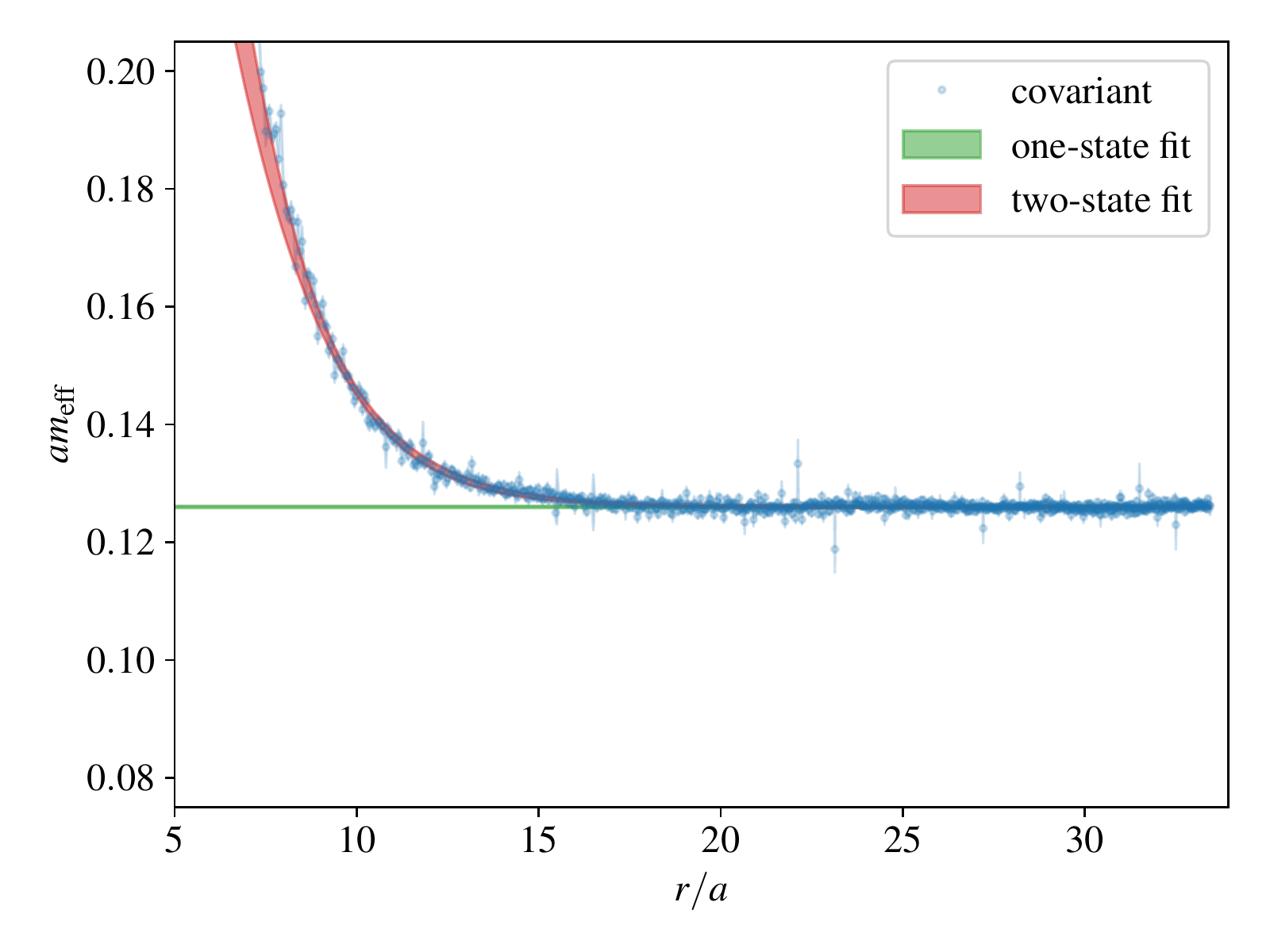}}
  \caption{%
    Effective mass of the $\mathring{C}_{PP}(r)$ correlator as a function of $r$ for master fields in set A (left plot) and set B (right plot).
    On top of the data points with master-field errors shown in blue, we show the results of a one-state fit in a green band and of a two-states fit in a red band.
    The thickness of the bands is the statistical error.
  }\label{fig:pion_mass}
\end{figure}

We employed the technique already studied in ref.~\cite{Ce:2021akh} to extract the pion mass $m_\pi$ from the long-distance behaviour in eq.~\eqref{eq:cPP} of the position-space correlator $\mathring{C}_{PP}(r)$.
In those proceedings the technique was applied to correlators computed with point sources on an ensemble of gauge field configurations with a $(\SI{6}{\fm})^3$ space volume, performing a standard error estimation.
Here we have a larger volume that allows us to use the grid of point sources as described in section~\ref{sec:grid_estimator} and estimate the error \emph{à la} master field, see section~\ref{sec:master_field_errors}.
On top of the same number of samples $n_{\text{point}}=4096$ for each configuration, we have $5$ configurations in set A and $2$ in set B.
This means that we have a larger statistics for the $L=96a$ master fields from which we expect a $\approx\num{1.58}$ reduction of the error.

The effective mass\footnote{
  See eq.~(\num{10}) in ref.~\cite{Ce:2021akh} for the definition of the effective mass of the radial correlator.
} of $\mathring{C}_{PP}(r)$ is shown in the two plots in figure~\ref{fig:pion_mass}.
For each set, two fits are performed: a \enquote{one-state} fit having $c_P$ and $m_\pi$ as free parameters, and a \enquote{two-states} one with an added \enquote{excited state} term $a_1(m_1/r)K_1(m_1/r)$ with two extra free parameters $a_1$ and $m_1>m_\pi$.
We choose appropriate values for the smaller $r$ of the correlator data that enter the fit, with different choices for one-state and two-states fits.
Instead, all the data up to largest available $r=r_{\text{max}}$ enter the fit, since we do not observe any boundary effect that constrains us otherwise.
The two fits on each set give compatible results and the corresponding effective mass is shown in figure~\ref{fig:pion_mass}.

From the one-state fits we obtain the results in table~\ref{tab:results}, which show a good agreement between the two sets.
Contrary to the expectation based on $n_{\text{cnfg}}$, the error is \SI{40}{\percent} smaller on set B.
A possible explanation for this fact is the $b_{\text{shift}}=12a$ of the samples of set A, halved with respect to set B, which can lead to a reduced effective number of samples due to stronger correlations in space.

\begin{figure}[t]
  \centering
    \scalebox{.45}{\includegraphics{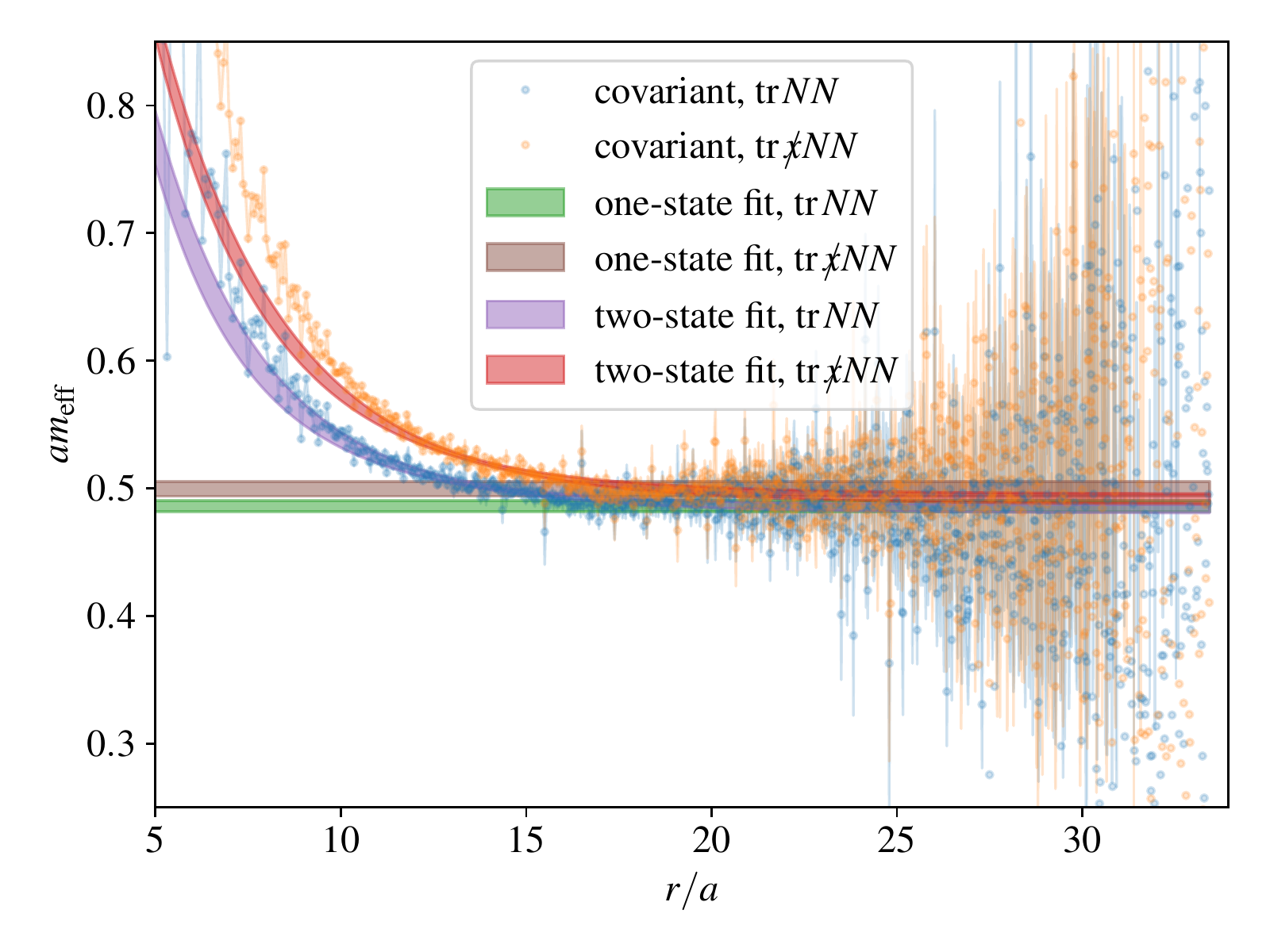}}%
    \scalebox{.45}{\includegraphics{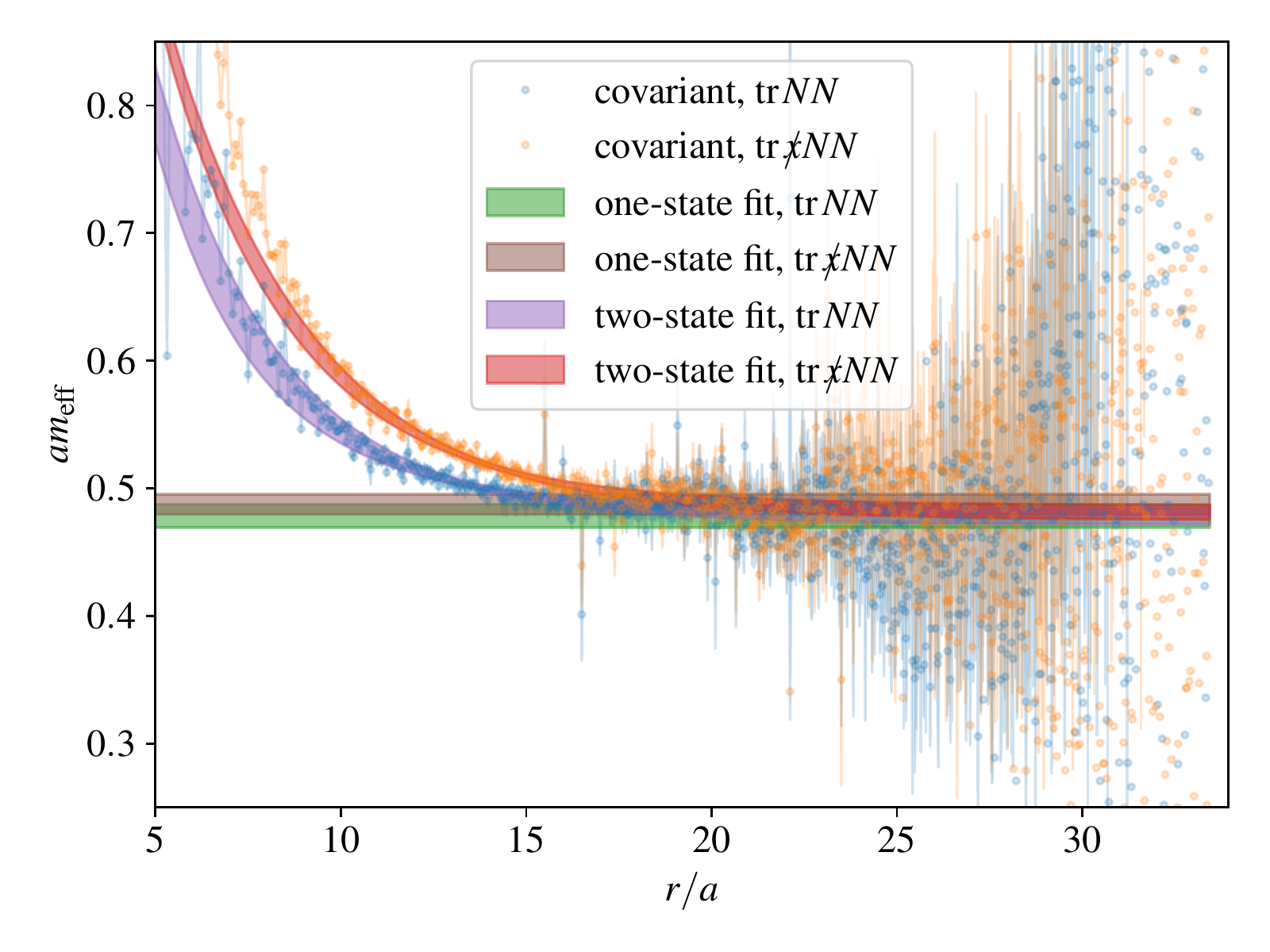}}
  \caption{%
      Effective mass of the $\mathring{C}^{(i)}_{NN}(r)$ correlators as a function of $r$ for master fields in set A (left plot) and set B (right plot), where $i=1$ corresponds to the $\tr NN$ contraction and $i=2$ to the $\tr\slashed{x}NN$ one.
    On top of the data points with master-field errors shown in blue and orange for $i=1$ and $2$ respectively, we show the results of a one-state fit in green and brown bands and of a two-states fit in red and purple bands.
    The thickness of the bands is the statistical error.
  }\label{fig:nucleon_mass}
\end{figure}

Similarly, we extract $m_N$ from the two contractions in eq.~\eqref{eq:cNN_ring} of the position-space nucleon correlator in eq.~\eqref{eq:cNN} as done in ref.~\cite{Ce:2021akh}, but employing the techniques of sections~\ref{sec:grid_estimator} and~\ref{sec:master_field_errors}.
The results in table~\ref{tab:results} are from the one-state fits to $\mathring{C}^{(1)}_{NN}(r)$ with the free parameters $c_N$ and $m_N$, and are compatible with the results of two-states fits with the replacement $\mathring{C}_{NN}(r)\to\mathring{C}_{NN}(r)[1+a_1(m_\pi/r)K_1(m_\pi r)]$ where $a_1$ is an extra free parameter and $m_\pi$ is fixed.
The fit to $\mathring{C}^{(2)}_{NN}(r)$ shows similar results, although with a slightly larger central value that can be attributed to different discretization effects.
The effective masses corresponding to data and fits are shown in figure~\ref{fig:nucleon_mass}.
In the case of $m_N$, we observe a larger error on set B, compatible with the lower statistics and showing no indication of correlation-in-space effects.

\begin{figure}[t]
  \centering
  \scalebox{.45}{\includegraphics{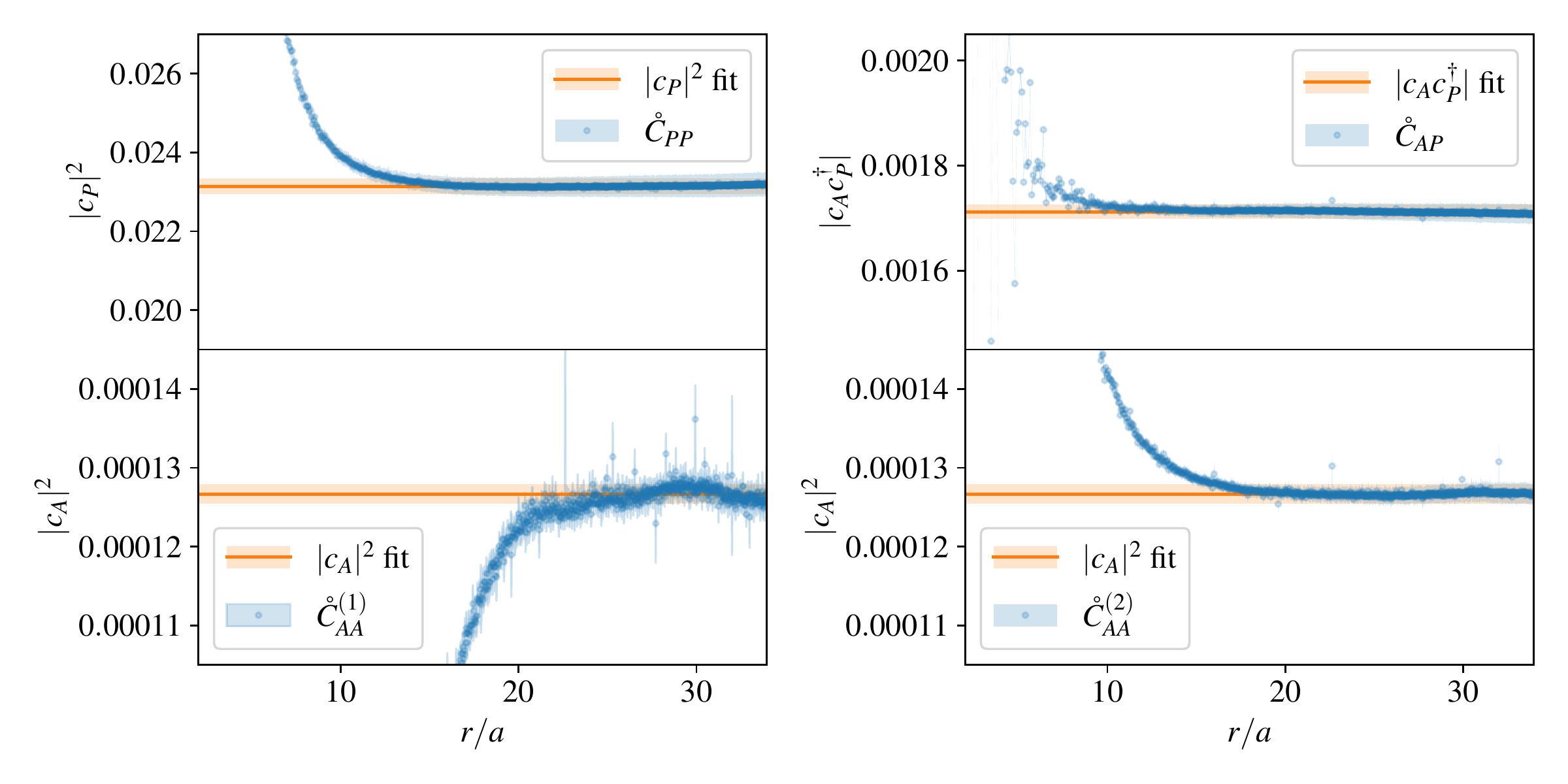}}\\
  \scalebox{.45}{\includegraphics{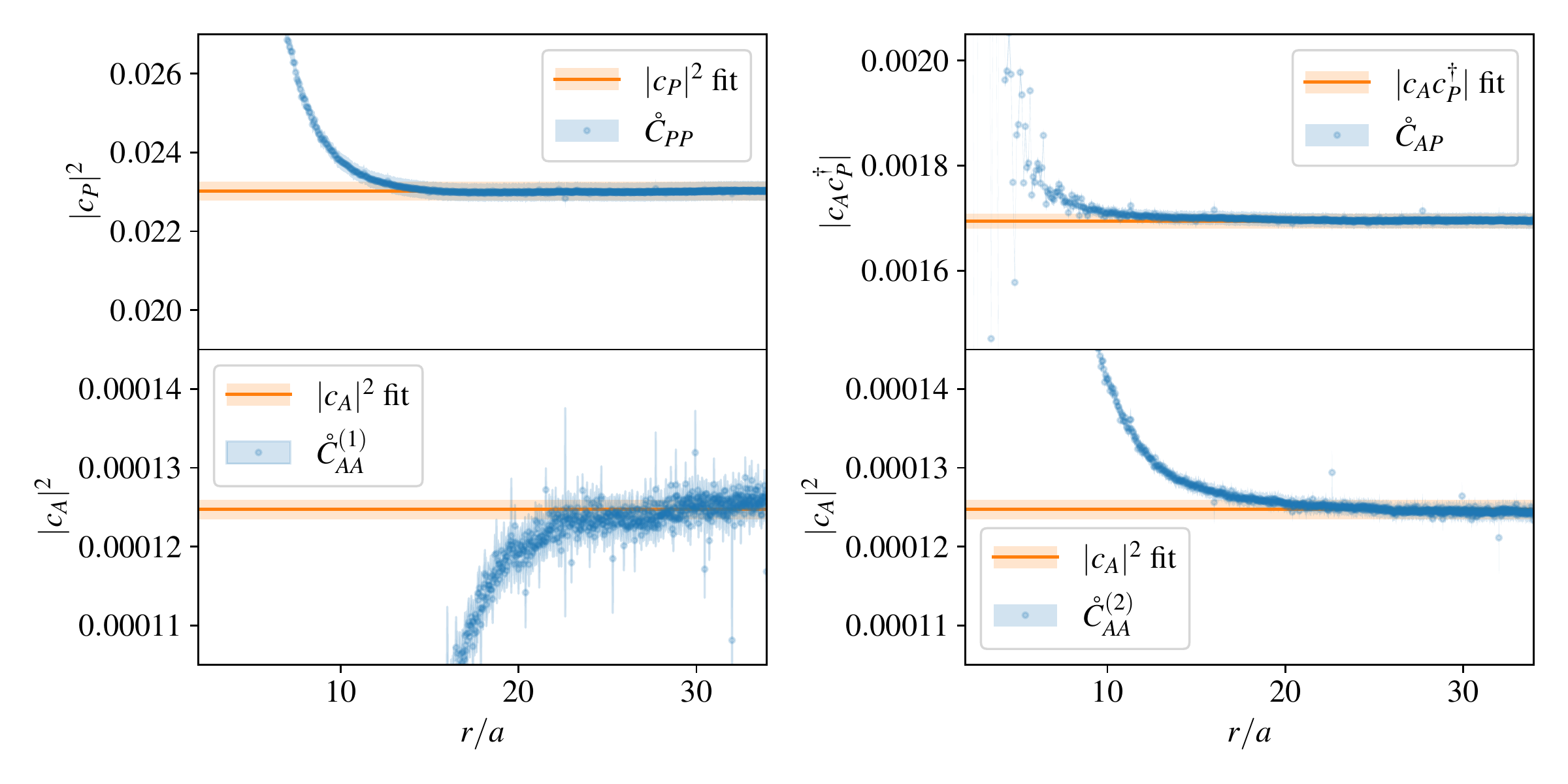}}
  \caption{%
    Plots of the ratio between correlator data and their fitted long-distance behaviours for master fields in set A (top row) and set B (bottom row).
    The amplitude in the denominator is set to one, so that the actual amplitude for each correlator is shown on the vertical axis.
    In each row, four plots are shown for $\mathring{C}_{PP}$, $\mathring{C}^{(1)}_{AA}$ (left column), $\mathring{C}_{AP}$ and $\mathring{C}^{(2)}_{AA}$ (right column), with the correlator data with master field errors shown in blue.
    The amplitude parameters of the corresponding fit function, which are functions of $c_P$ and $c_A$, are shown in an orange horizontal line with a pale orange error band.
  }\label{fig:decay_constant}
\end{figure}

We also extract the pion decay constant $f_\pi^{\text{bare}}$, where the \emph{bare} indicates that we do not include the axial-current renormalization factor, from a combined fit of the four correlators $\mathring{C}_{PP}$, $\mathring{C}_{AP}$, $\mathring{C}^{(1)}_{AA}$ and $\mathring{C}^{(2)}_{AA}$.
As fit function we employ the long-distance behaviours derived from eqs.~\ref{eq:position_space_corr}, which depends on the free parameters $c_P$, $c_A$ and $m_\pi$.
As shown from the plots of the ratio between data and fit functions in figure~\ref{fig:decay_constant}, $\mathring{C}_{AP}$ approaches the asymptotic behaviour at a smaller value of $r$, followed by $\mathring{C}_{PP}$ and $\mathring{C}^{(2)}_{AA}$.
$\mathring{C}^{(1)}_{AA}$ converges to the asymptotic behaviour at a much larger $r$, with the ratio being initially negative and changing sign around $r\approx 14a$.
The values of $m_\pi$ obtained from these combined fits are consistent with the previous fits to only the $\mathring{C}_{PP}$ correlators.
The decay constant is then given by $f_\pi^{\text{bare}}=c_A/m_\pi$ and shown in table~\ref{tab:results}.
Like in the case of $m_N$, the values on set A and B are compatible, with a slightly larger error for set B that is consistent with the lower number of master field configurations.

\enlargethispage*{\baselineskip}
\section{Conclusions}

We have shown that position-space correlators can be used to extract hadron masses and decay constants with short-distance and cut-off effects under control.
Crucially, the statistical error can be estimated \emph{à la} master field, obtaining an efficient scaling of the computational effort with the increased volume.

In this work we studied sphere-averaged radial correlators, but potentially more information is encoded in correlators as function of four-dimensional coordinates.
This requires understanding effects that break rotational symmetry at finite lattice spacing and is an interesting topic for further studies.

Position-space methods find applications in computations of quantities that go beyond the simple hadronic quantities considered here, such as for example the hadronic vacuum polarisation contribution to the anomalous magnetic moment of the muon~\cite{Meyer:2017hjv,Ce:2018ziv}, including the so-called window contribution~\cite{Chao:2022ycy}.
The estimators presented here provide a straightforward path to the computation of this quantities in the master-field paradigm.


{\small\noindent\textbf{Acknowledgements:}
The research of MB is funded through the MUR program for young researchers \enquote{Rita Levi Montalcini}.
AF acknowledges support by the Ministry of Science and Technology Taiwan (MOST)
under grant 111-2112-M-A49-018-MY2.
JRG acknowledges support from the Simons Foundation through the Simons Bridge for Postdoctoral Fellowships scheme.
MTH is supported by UKRI Future Leader Fellowship MR/T019956/1 and in part by UK STFC grant ST/P000630/1.
This work was performed using the DiRAC Data Intensive service at Leicester, operated by the University of Leicester IT Services, which forms part of the STFC DiRAC HPC Facility (\url{www.dirac.ac.uk}).
The equipment was funded by BEIS capital funding via STFC capital grants ST/K000373/1 and ST/R002363/1 and STFC DiRAC Operations grant ST/R001014/1.
DiRAC is part of the National e-Infrastructure.
We acknowledge PRACE for awarding us access to SuperMUC-NG at GCS@LRZ, Germany, where some computations were performed
Many simulations were performed on a dedicated HPC cluster at CERN.
We gratefully acknowledge the computer resources and the technical support provided by these institutions.
}

\setlength{\bibsep}{0pt plus 0.3ex}
\bibliography{./biblio_bibertool.bib}

\end{document}